\documentclass[sigconf]{acmart}

\usepackage{booktabs}
\usepackage{array} 
\usepackage{multirow} 
\usepackage{tabularx}
\usepackage{geometry} 
\usepackage{natbib}

\AtBeginDocument{%
  }

\setcopyright{acmlicensed}
\copyrightyear{2024}
\acmYear{2024}
\acmDOI{XXXXXXX.XXXXXXX}

\acmConference[Conference acronym 'XX]{Make sure to enter the correct
  conference title from your rights confirmation emai}{June 03--05,
  2018}{Woodstock, NY}
\acmISBN{978-1-4503-XXXX-X/18/06}

\author{Zhichao Geng}
\email{zhichaog@amazon.com}
\affiliation{%
  \institution{Amazon}
  \city{Shanghai}
  \country{China}}

\author{Yiwen Wang}
\email{wangyiwe@amazon.com}
\affiliation{%
  \institution{Amazon}
  \city{Shanghai}
  \country{China}}

\author{Dongyu Ru}
\email{rudongyu@amazon.com}
\affiliation{%
  \institution{Amazon}
  \city{Shanghai}
  \country{China}}

\author{Yang Yang}
\email{yych@amazon.com}
\affiliation{%
  \institution{Amazon}
  \city{Shanghai}
  \country{China}}

\begin{document}

\title{Towards Competitive Search Relevance For Inference-Free Learned Sparse Retrievers}

\begin{abstract}
  Learned sparse retrieval, which can efficiently perform retrieval through mature inverted-index engines, has garnered growing attention in recent years. Particularly, the inference-free sparse retrievers are attractive as they eliminate online model inference in the retrieval phase thereby avoids huge computational cost, offering reasonable throughput and latency.
  However, even the state-of-the-art (SOTA) inference-free sparse models lag far behind in terms of search relevance when compared to both sparse and dense siamese models. Towards competitive search relevance for inference-free sparse retrievers, we argue that they deserve dedicated training methods other than using same ones with siamese encoders. In this paper, we propose two different approaches for performance improvement. 
  First, we propose an IDF-aware penalty for the matching function that suppresses the contribution of low-IDF tokens and increases the model’s focus on informative terms.
  Moreover, we propose a heterogeneous ensemble knowledge distillation framework that combines siamese dense and sparse retrievers to generate supervisory signals during the pre-training phase. The ensemble framework of dense and sparse retriever capitalizes on their strengths respectively, providing a strong upper bound for knowledge distillation. 
  To concur the diverse feedback from heterogeneous supervisors, we normalize and then aggregate the outputs of the teacher models to eliminate score scale differences.
  On the BEIR benchmark, our model outperforms existing SOTA inference-free sparse model by \textbf{3.3 NDCG@10 score}.  It exhibits search relevance comparable to siamese sparse retrievers and client-side latency only \textbf{1.1x that of BM25}.
\end{abstract}



\received{20 February 2007}
\received[revised]{12 March 2009}
\received[accepted]{5 June 2009}

\maketitle

\section{Introduction}
Information retrieval(IR) and question answering(QA) are fundamental tasks in the realm of information processing, widely employed in various web applications. 
Lexical-based algorithms such as TF-IDF and BM25 were once the dominant approach. These algorithms utilize inverted indexes, which was proven to be efficient. However, due to issues such as vocabulary mismatch~\cite{zhao2010term} and the lack of contextual information, their semantic retrieval capabilities are limited. In contrast, siamese dense retrievers have overcome the limitations of traditional lexical-based methods and have become the mainstream approach for semantic retrieval~\cite{reimers-2019-sentence-bert}. Nonetheless, the ANN algorithm requires a substantial amount of memory leading to a significant trade-off between search relevance and resource consumption~\cite{malkov2018efficient,jegou2010product}. The interpretability of dense retrievers is also questioned. Recent years, learned sparse retrieval is proposed to address this obstacle and gained increasing attention~\cite{dai2020context,formal2021splade, formal2021splade2}. This approach predicts token weights based on their semantics with context information. It expands token set with generative models~\cite{nogueira2019doc2query, nogueira2019document} or masked language model heads~\cite{bai2020sparterm,zhao2021sparta, formal2021splade, formal2021splade2,formal2022distillation, lassance2022efficiency,macavaney2020expansion,lassance2024splade}, thereby addressing the vocabulary mismatch problem. Since sparse embeddings can be integrated with inverted indexes, the retrieval process of learned sparse models is highly efficient without compromising recall. Moreover, learned sparse retrieval offers better interpretability because the contribution of each token can be intuitively understood by human.

Among the learned sparse models, the inference-free architecture is particularly attractive to search applications. This architecture degenerates online model inference for query encoding into simple tokenization, significantly reducing end-to-end search latency and the associated model deployment costs. 

Early work like DEEPCT~\cite{dai2020context} and Doc2Query~\cite{nogueira2019doc2query} attempted to associate additional information with the original documents. However, search relevance could not be trained in an end-to-end manner.
Proposed by \citet{formal2021splade2}, the \textit{SPLADE-doc} architecture has achieved SOTA performance among inference-free retrievers. It predicts the token weights and expands the tokens with similar semantics. The search relevance and sparsity are tuned via end-to-end training.
However, even the latest SOTA inference-free model, \textit{SPLADE-v3-Doc}~\cite{lassance2024splade}, exhibits a significant gap in search relevance when compared with siamese sparse retrievers. On the BEIR benchmark, the average NDCG@10 score of \textit{SPLADE-v3-Doc} is 4.7 lower than siamese sparse retrievers of the same size and training method. This disparity hinders its application in actual production environments.

In this paper, we focus on improving the search relevance of the inference-free sparse retriever through more effective training methodologies. 
The first challenge lies in the uniform penalty applied to all tokens by the FLOPS regularization~\cite{paria2020minimizing}. 
We argue that the uniform penalty unintentionally suppresses rare but important terms. To address this, we propose a penalty strategy that adjusts token importance based on their inverse document frequency (IDF). Specifically, we integrate IDF weights into the scoring mechanism, which encourages the model to assign higher relevance scores to informative low-frequency tokens. This, in turn, guides the optimization to preserve such tokens during training, improving relevance and maintaining overall sparsity.
Through experiments, we demonstrate that IDF-aware penalty effectively improves the search relevance of the inference-free sparse retriever, and we discover that it effectively reduces the average FLOPS number for the retrieval process.

Subsequently, the pre-training phase is also explored in this paper. We argue that although the commonly used contrastive InfoNCE loss~\cite{chen2020simple} is able to enhance the alignment and uniformity of representations~\cite{wang2020understanding}, these two targets are not applicable in inference-free models. Because for inference-free models, all semantics are only encoded at the model-side. And the search relevance can not be improved by aligning the document representation with the bag-of-word query representation. In contrast, knowledge distillation presents a more suitable approach for the training~\cite{formal2021splade2}. \citet{hofstatter2020improving} proposed to train dense retrievers by conducting knowledge distillation from cross-encoder rerankers, and \citet{formal2022distillation, formal2021splade2} applied this method to the fine-tuning of sparse retrievers. However, for large-scale pre-training datasets, the inference workload of the teacher model can reach 10 times or even larger. And the inference cost of cross-encoders is impractical at these settings, especially when in-batch negatives are utilized. In this paper, we propose to build a strong teacher model by assembling siamese dense retrievers and siamese sparse retrievers. Siamese retrievers have a heterogeneous and superior architecture compared with the inference-free architecture. And their inference cost is applicable for large-scale pre-training. Moreover, the ensemble of dense and sparse retrievers can further enhance the upper bound of knowledge distillation, enlarging the space for performance improvement of our model. During the assembling process, we normalize the scores for heterogeneous retrievers. This prevents one retriever from dominating the assembled result, further balancing the contribution of teacher models. 

We conduct experiments on 13 public datasets from the BEIR benchmark, and our model outperforms the existing SOTA inference-free sparse model by 3.3 average NDCG@10 scores. Its performance even surpasses many strong siamese retrievers. Our contributions can be summarized as follows: 
    (1) We propose IDF-aware penalty, which effectively improves the search relevance and efficiency of inference-free sparse models. 
    (2) We explore how to effectively pre-train inference-free sparse models and propose the ensemble teacher model of heterogeneous siamese models, which has reasonable inference costs and strong performance. 
    (3) The zero-shot performance of our model outperforms the SOTA inference-free retriever by \textbf{3.3 NDCG@10 score}. It also surpasses strong siamese retrievers including \textit{SPLADE-v3-DistilBERT} and \textit{ColBERTv2}. While its client-side latency is only \textbf{1.1x that of BM25}.

\section{Related Work}
\subsection{Dense Retrieval}
Recent years, the use of language models to generate dense embeddings for text representations has become prevalent in QA and IR~\cite{reimers-2019-sentence-bert, conneau2017supervised, karpukhin2020dense, qu2021rocketqa}. Continuous efforts have been made to improve the training methodologies for dense retrieval models, such as negative sampling~\cite{qu2021rocketqa,zhan2021optimizing} and knowledge distillation~\cite{hofstatter2020improving, hofstatter2021efficiently, lin2021batch}. To enhance the generalization capability of dense retrieval models, numerous studies explore pre-training techniques on text embeddings. Some works~\cite{gao2021condenser, gao2022unsupervised, xiao2022retromae} design auxiliary tasks to enrich the dense embeddings from models, while another tributary of previous work pre-train the model directly on constructed text pairs~\cite{li2023towards, chen2024bge, wang2022text}, including unsupervised and weakly supervised data.

Knowledge distillation~\cite{gou2021knowledge} utilizes the soft labels from teacher models to facilitate a more effective training process for student models so as to improve the accuracy. Researchers strive to select teacher models with inherent advantages, such as larger parameters size or superior architectures, to ensure the best possible knowledge transfer. \citet{hofstatter2020improving} proposes using a cross-encoder reranker as the teacher model for the siamese dense retriever during fine-tuning. However, in the context of pre-training, the significantly larger data volume makes the use of cross-encoders prohibitively expensive.

Pre-training is a widely adopted technique to enhance the accuracy and generalization capabilities of dense retrievers. The contrastive InfoNCE loss is applied to massive amounts of unsupervised or weakly-supervised data. Previous work~\cite{wang2020understanding} illustrates that the contrastive InfoNCE loss improves the dense representation from the perspectives of alignment and uniformity. However, for the inference-free architecture, the asymmetric sparse representation is unaware of the query distribution, and the token weight is simply an importance measure for that token. Consistent with our experiments, pre-training with the InfoNCE loss does not improve the model as expected on dense retrievers.

The challenges associated with knowledge distillation and pre-training methodologies limit the application of these techniques to inference-free sparse retrievers.

\subsection{Sparse Retrieval}
Learned sparse retrievers have gained increasing attention due to their ability to perform semantic search while retaining the advantages of traditional lexical-based retrieval methods. DEEPCT~\cite{dai2020context} employs the BERT model to fit the token weights computed by heuristic rules. It modifies the term frequency, thereby influencing the match score in the BM25 algorithm. However, DEEPCT does not address the vocabulary mismatch issue inherent in lexical-based retrieval. Doc2query~\cite{nogueira2019document} and docTTTTTquery~\cite{nogueira2019doc2query} tackle this issue by using generative models to predict potential queries for the original document. These queries are indexed together with the original document, enabling the matching of tokens not present in the document. These methods represent a preliminary exploration of inference-free learned sparse retrievers. Nevertheless, these models are unable to apply ranking loss in an end-to-end manner, thereby limiting their performance.

Following these research studies, more works have been proposed to train sparse retrievers in an end-to-end manner. Ranking losses such as infoNCE which are already widely used in training dense retrievers are discovered to be also eligible for sparse retrievers. Methods such as SparTerm~\cite{bai2020sparterm}, EPIC~\cite{macavaney2020expansion}, and SPARTA~\cite{zhao2021sparta} predict token weights based on the estimation of token importance and then apply various sparsification strategies, such as top-k pooling or learned gating, to obtain sparse embeddings. The match score between the query and document is obtained by taking the inner product of their sparse embeddings. Since the output space is identical to the vocabulary space, tokens that do not appear in the document can also be expanded in the representation. However, these methods are not of an effective end-to-end optimization manner and did not combine ranking loss and sparsity simultaneously.

The SPLADE-series models~\cite{formal2021splade,formal2021splade2,formal2022distillation,lassance2024splade} overcome this obstacle by introducing the FLOPS regularizer to the loss function, enabling an end-to-end integration of ranking loss and sparsity. Inspired by \citet{hofstatter2020improving}, they employ hard negatives and knowledge distillation from cross-encoders to improve model performance. Among the SPLADE-series models, \textit{SPLADE-doc}~\cite{formal2021splade2} performs model inference solely on documents and sums up the weights for all matched tokens as the match score. It eliminates the need for model inference during retrieval, and the model training is completely end-to-end. However, even SPLADE-doc offers SOTA performance among inference-free models, there is still a significant gap persists between it and siamese dense/sparse retrievers in terms of search relevance.

\section{Preliminary}
Our work is built upon the \textit{SPLADE-doc-distill}\footnote{The original paper~\cite{formal2021splade2} proposes the \textit{SPLADE-doc} architecture and knowledge distillation training method. We incorporate the knowledge distill on \textit{SPLADE-doc} and call it \textit{SPLADE-doc-distill}. We implement this baseline in the experiment session.} model. In this section, we’ll introduce the details of the baseline method. 

\subsection{Ranking supervision.}
To capture semantic relevance, SPLADE-doc-distill employs knowledge distillation in which the student model learns from the teacher models. Let $\mathbf{s}_{\text{tea}}$ and $\mathbf{s}_{\text{stu}}$ denote the teacher and student scores across documents. The ranking loss is defined as the KL divergence between their softmax distributions:
\begin{equation}
\begin{aligned}
\mathcal{L}_{\text{rank}} = \text{KL} \big( \,
     \text{softmax}(\mathbf{s}_{\text{tea}}) 
    \|\,  \text{softmax}(\mathbf{s}_{\text{stu}}) \,
\big),
\end{aligned}
\label{eq:ranking_loss}
\end{equation}
where $\mathbf{s}_{\text{stu}} = [s(q, d_1), \dots, s(q, d_n)]$ denotes the scores computed by the student model between the query $q$ and each candidate document $d_i$, and

\begin{equation}
s(q, d_i) = \sum_{t \in \mathcal{V}} q_t \cdot d_{i,t},
\label{eq:match_score}
\end{equation}
with $q_t \in \{0,1\}$ indicating the presence of token $t$ in query $q$, and $d_{i,t}$ representing the activation value of token $t$ in document $d_i$.

\subsection{Sparsity regularization.}
To promote sparsity in the document vectors, it applies a FLOPS regularizer~\cite{paria2020minimizing} that penalizes high average token activations. For a batch of $N$ documents, let $w_j^{(d_i)}$ be the activation of token $j$ in document $d_i$. The FLOPS loss is defined as the sum of squared average activations:
\begin{equation}
\mathcal{L}_{\text{FLOPS}} = \sum_{j \in \mathcal{V}} \left( \frac{1}{N} \sum_{i=1}^{N} w_j^{(d_i)} \right)^2.
\label{eq:FLOPS_loss}
\end{equation}
The final training loss jointly optimizes relevance alignment and representation sparsity:
\begin{equation}
\mathcal{L} = \mathcal{L}_{\text{rank}} + \lambda_d \cdot \mathcal{L}_{\text{FLOPS}},
\label{eq:total_loss}
\end{equation}
where $\lambda_d$ is a hyperparameter controlling the trade-off between matching accuracy and sparsity.

\section{Method}
\subsection{IDF-aware Penalty}
\paragraph{Observation.} To better understand how existing sparsity regularization behaves in retrieval models, we analyze token-level activation and FLOPs penalty scores using a standard SPLADE-v3-doc model~\cite{lassance2024splade} on scidocs dataset~\cite{thakur2021beir}. 
\begin{figure}
    \centering
    \includegraphics[width=\linewidth]{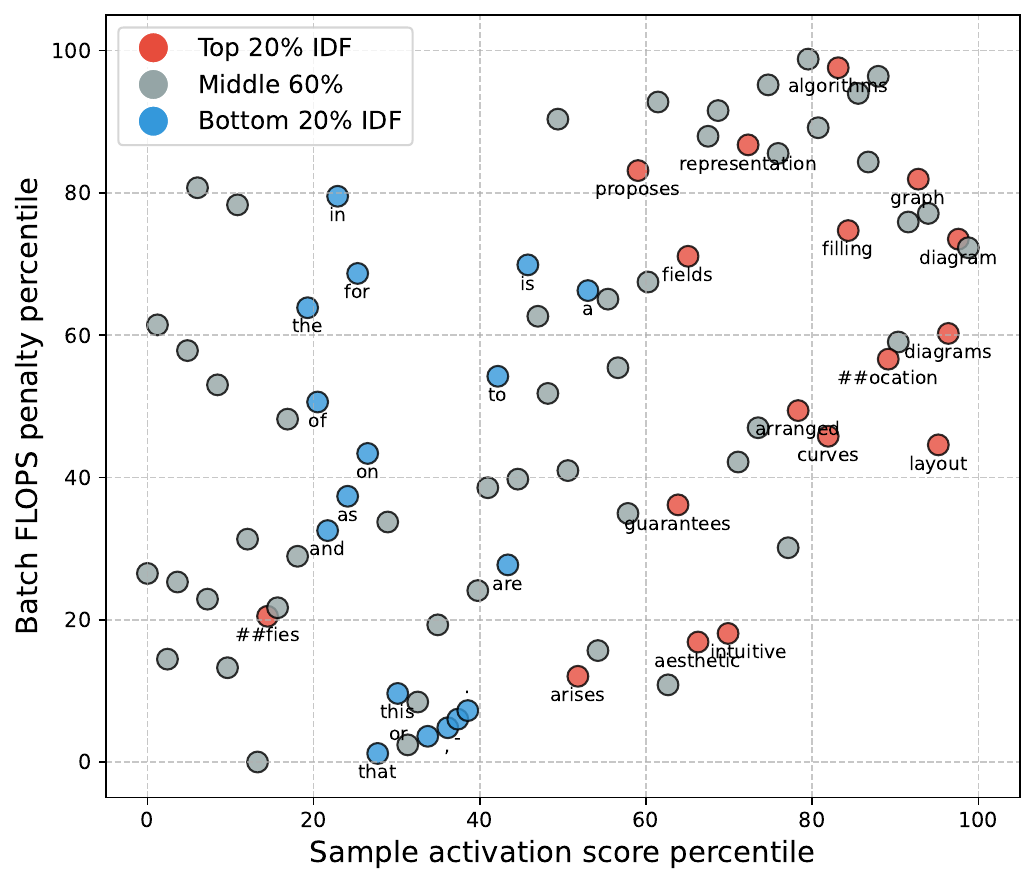}
    \caption{Token-level activation vs. FLOPs penalty in a random batch encoded by SPLADE-v3-doc. The sparsity regularization under-penalizes trivial tokens while suppresses informative ones.}
    \label{fig:token_flops_scatter}
\end{figure} 
As illustrated in Figure~\ref{fig:token_flops_scatter}, the distribution shows that many trivial tokens—such as "that" and "this"—receive low FLOPs penalties despite being activated, while more meaningful tokens like "algorithms" or "graph" are penalized disproportionately. This indicates that the standard FLOPs regularization lacks semantic awareness, treating all tokens uniformly regardless of their informational value.

We further observe that tokens with higher activation scores tend to have higher IDF values, suggesting that IDF serves as an effective indicator of token importance within document representations. Motivated by this insight, we incorporate IDF as a guiding prior in the loss function to address the existing limitation.

\paragraph{IDF-aware Penalty.} Building on our observation, we introduce an IDF-aware Penalty to SPLADE-doc-distill by modifying the ranking objective. Specifically, we define an IDF-weighted ranking loss, $\mathcal{L}_{\text{rank-idf}}$, as a refinement of the original ranking loss $\mathcal{L}_{\text{rank}}$ in Eq.~\ref{eq:ranking_loss}, in which the matching score in Eq.~\ref{eq:match_score} is redefined as:
\begin{equation}
s(q, d_i) = \sum_{t \in \mathcal{V}} \text{idf}(t) \cdot q_t \cdot d_{i,t},
\label{eq:idf_weighted_score}
\end{equation}
with $\text{idf}(t)$ denoting the IDF value of token $t$. This adjustment alters the gradient dynamics during training. The final training objective combines $\mathcal{L}_{\text{rank-idf}}$ with a FLOPS-based regularization term: 
\begin{equation}
\mathcal{L} = \mathcal{L}_{\text{rank-idf}} + \lambda \cdot \mathcal{L}_{\text{FLOPS}}.
\end{equation}
For each token $t$, the gradient of the total loss with respect to its document-side activation $d_{i,t}$ can be decomposed as:
\begin{equation}
\frac{\partial \mathcal{L}}{\partial d_{i,t}} =
\frac{\partial \mathcal{L}_{\text{rank-idf}}}{\partial d_{i,t}} + \lambda \cdot \frac{\partial \mathcal{L}_{\text{FLOPS}}}{\partial d_{i,t}}.
\end{equation}
The first term, originating from the ranking loss, is proportional to the token’s IDF value:
\[
\frac{\partial \mathcal{L}_{\text{rank-idf}}}{\partial d_{i,t}} \propto \text{idf}(t) \cdot q_t \cdot (\text{softmax}_\text{stu}-\text{softmax}_\text{tea}).
\]
while the FLOPS regularization imposes a uniform penalty (e.g., $2d_{i,t}$).

This composition reveals an important insight: For tokens with high $\text{idf}(t)$, the gradient from the ranking loss dominates, encouraging the model to preserve their activations. In contrast, tokens with low IDF values receive weaker ranking gradients and are thus more susceptible to sparsification by the FLOPS regularization. As a result, the model learns to retain informative tokens while effectively penalizing unimportant ones, leading to efficient retrieval with minimal performance degradation.

\begin{table*}[htbp]
\centering
\caption{The list of datasets used in pre-training.}
\label{tab:datasets}
\small
\begin{tabular}{cccc}
\toprule
\textbf{Dataset} & \textbf{Type} & \textbf{Number of Training Tuples} & \textbf{Remaining Queries After Filtering} \\
\midrule
S2ORC~\cite{lo2020s2orc} & (Title, Abstract) pairs & 41,769,185 & 500,000 \\
WikiAnswers~\cite{fader2014open} & duplicate questions & 77,427,422 & 1,000,000 \\
GOOAQ~\cite{khashabi2021gooaq} & (Question, Answer) pairs & 3,012,496 & 2,274,901 \\
SearchQA & Question + Top5 text snippets & 117,220 & 116,933 \\
Eli5~\cite{fan2019eli5} & (Question, Answer) pairs & 325,475 & 168,652 \\
WikiHow~\cite{koupaee2018wikihow} & (Summary, Text) pairs & 128,542 & 111,987 \\
SQuAD~\cite{rajpurkar2018know} & (Question, Answer\_Passage) pairs & 87,599 & 84,505 \\
Stack Exchange & (Title, Title) pairs of duplicate questions & 304,525 & 132,490 \\
 & (Body, Body) pairs of duplicate questions & 250,519 & 98,266 \\
 & (Title+Body, Title+Body) pairs of duplicate questions & 250,460 & 117,839 \\
Yahoo Answers~\cite{zhang2015character} & (Title, Answer) pairs & 1,198,260 & 361,312 \\
 & (Question, Answer) pairs & 681,164 & 139,584 \\
 & (Title, Question) pairs & 659,896 & 252,823 \\
\bottomrule
\end{tabular}
\end{table*}

\subsection{Ensemble Heterogeneous Knowledge Distillation}
\label{sec:ensemble_hete}

Pre-training is a widely adopted technique to improve the performance and generalization of retrieval models. To further boost the search relevance, we pre-train the model on an extensive corpus encompassing both unsupervised and weakly-supervised datasets. Subsequently, we fine-tune the model on a high-quality labeled dataset, specifically the MS MARCO dataset. To construct an effective optimization objective, we employ knowledge distillation techniques.
The primary challenge lies in generating efficient supervisory signals for the large-scale pre-training data. For an input batch containing $N$ queires and $M$ documents, the cross-encoders need to inference at $O(MN)$ complexity. And the cost becomes impractical as data scales.

\begin{figure}[ht]
\caption{The procedure for pre-training with ensemble heterogeneous knowledge distillation.}
  \includegraphics[width=\columnwidth]{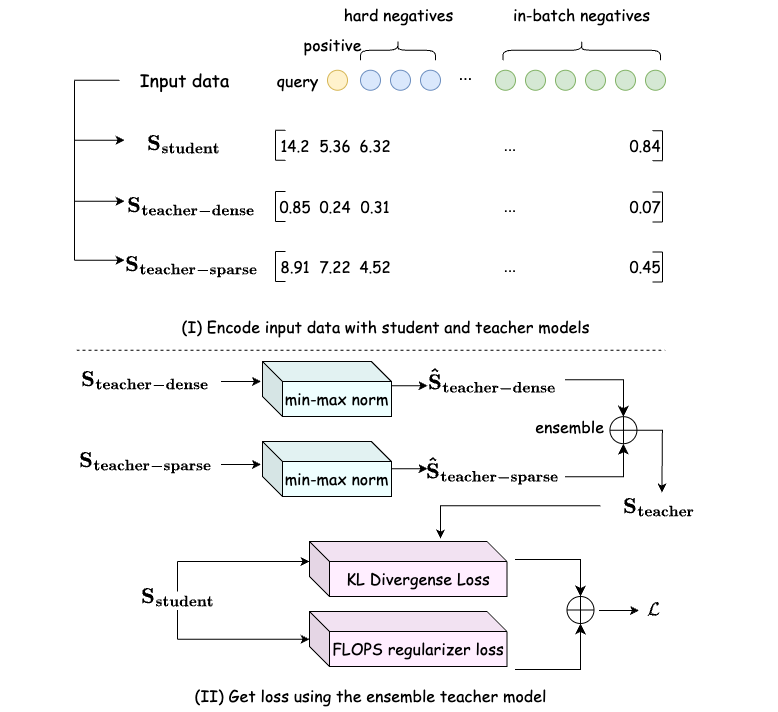}
  \label{fig:workflow}
\end{figure}

In this paper, we introduce a novel technique that leverages an ensemble of heterogeneous models as the teacher model for knowledge distillation on large-scale data, the assembling procedure is illustrated in Figure \ref{fig:workflow}. Siamese dense and sparse retrievers are combined to generate supervisory signals for the input data, and we employ the KL Divergence loss function to transfer knowledge to the student model.

\textbf{Heterogeneous teacher models.} The cross-encoders are used by previous work~\cite{formal2022distillation} to provide supervision signals. However, their inference cost is impractical to be applied on large-scale pre-training. In contrast to inference-free retrievers, siamese retrievers possess a superior architecture and significantly lower inference costs compared with cross-encoders. To compensate the accuracy drop compared to cross-encoders, we propose an ensemble approach that combines dense and sparse retrievers as the teacher model. 
Sparse and dense teachers emit different document recall process, forming a heterogeneous distillation framework.
Research studies~\cite{chen2024bge, buhlmann2012bagging} and industrial practices\footnote{\url{https://opensearch.org/blog/semantic-science-benchmarks/}} demonstrate that combining these predictors results in a significantly more robust retriever. Consequently, the teacher model comprising an ensemble of dense and sparse retrievers is both efficient and accurate simultaneously. But there are still challenges for combining dense and sparse retrievers, as their match scores have disparate scaling. If not addressed properly, the retriever with a larger score scale may dominate the combined results, leading to biased or skewed outputs. For each retriever the scores are normalized before integration. We employ min-max normalization to scale all scores to the range of [0,1]. Subsequently, we combine the two retrievers by calculating their arithmetic mean and get the weighted sum. We then multiply the combined scores with a constant $S$ to scale them back for knowledge distillation. These steps ensure a balanced contribution from both sides, which are crucial for combining diverse retrieval models. The aforementioned process can be represented by the following formula:
\begin{equation}
    \hat{s_i^j}=\frac{s_i^j-min(s^j)}{max(s^j)-min(s^j)} 
\end{equation}
\begin{equation}
    \hat{s}= S\cdot \sum_{j}w^js^j
\end{equation}
where $s_i^j$ are the match scores from model $j$ on document $i$, $w_j$ is the weight for model $j$, and $\hat{s}$ are the final supervisory signals. In our experiments, we use equal weight for the dense and sparse teacher model.

We use the KL Divergence to compute the loss for the ensemble scores and the output of our inference-free sparse model. The FLOPS regularizer is also applied during the pre-training phase. We utilize a small coefficient $\lambda_d$ for the FLOPS regularizer. The first reason is that for the inference-free sparse model, the involved tokens are limited, and we aim to avoid omitting any token during the pre-training phase. Secondly, we need to search the optimal hyperparameter $\lambda_d$ to strike a balance between search relevance and retrieval cost through multiple experiments. It will be more efficient to conduct these experiments solely for fine-tuning.

\textbf{Data preparation.} Regarding the pre-training phase, we use a subset of training data collected by Sentence Transformers.
\footnote{\url{https://huggingface.co/sentence-transformers/all-MiniLM-L6-v2\#training-data}} 
The pre-training datasets were presented in the form of pairs, such as (Question, Answer), (Title, Content), and duplicate question/answer sets collected by the content providers. They are constructed either through automatic rules or human annotation, covering multiple domains. To harness the full potential of knowledge distillation, self-mined hard samples are used during training. We first train an inference-free sparse retriever without the pre-training phase as the miner model. Subsequently, for each query we use this model to mine the top $M$ relevant documents from the full documents collection of the source dataset. In each pre-training step, we commence by randomly selecting a dataset, followed by the random sampling of $N$ queries and their corresponding hard samples.
Inspired by \citet{wang2022text}, we use a consistency-based filtering approach to retain only those training samples where the labeled positive document is ranked among the top-$k$ retrieved documents.

Subsequent to the pre-training phase, we fine-tune the model on the MS MARCO dataset. Following~\citet{formal2021splade2}, cross-encoders are utilized as teacher models during fine-tuning. We ensemble the cross-encoders with dense and sparse teacher models used in pre-training. At fine-tuning stage, we accomplish the final sparsification of the representation.

\section{Experiments}
\label{sec:exp}
\subsection{Settings}
\subsubsection{Training Data}
For the pre-training phase, we utilized a subset of the datasets collected by the Sentence Transformers project. Detailed datasets are listed in Table \ref{tab:datasets}. Following the last paragraph in Section \ref{sec:ensemble_hete}, samples where the positive document is not ranked among the top 10 results are filtered out. In each training step, we take the positive document and 7 hard negative documents for every query. For the S2ORC and WikiAnswers datasets, we only sample a portion of them to prevent the large sample size from dominating the pre-training dataset. Ultimately, there are 5359292 queries and their corresponding hard negative documents.
For the fine-tuning phase, we utilize the MS MARCO passage ranking dataset. This dataset comprises 8,841,823 passages and 502,939 queries in the training set. For each query, we sample the top 100 hard negative documents to facilitate knowledge distillation.

\subsubsection{Model Training}
Co-Condenser~\cite{gao2022unsupervised}\footnote{\url{https://huggingface.co/Luyu/co-condenser-marco}} is hired as the backbone, which is of the same size with \textit{BERT-base} model. The IDF values for tokens are calculated using the documents of MS MARCO dataset. If a token is not present in the dataset, its value is set to 1.  The query IDF representation remains frozen throughout the training and evaluation processes. 

For the teacher models of the pre-training phase, we employ SOTA dense and sparse retrievers , namely \textit{gte-large-en-v1.5}\footnote{\url{https://huggingface.co/Alibaba-NLP/gte-large-en-v1.5}} and \textit{opensearch-neural-sparse-encoding-v1}\footnote{\url{https://huggingface.co/opensearch-project/opensearch-neural-sparse-encoding-v1}}. The total number of pre-training steps is 150,000. For each step, we sample 48 queries and 8 hard samples(1 positive document and 7 hard negative documents) for every query. We set the learning rate to 5e-5 and the FLOPS $\lambda_d$ to 1e-7. The max input length is set to 128. The constant $S$ is set to 10 for scaling back the assembled scores.
In the fine-tuning phase, we utilize an ensemble teacher model comprising the two siamese retrievers used in the pre-training phase, as well as two cross-encoder re-rankers\footnote{We’re using \url{https://huggingface.co/castorini/monot5-3b-msmarco-10k} and \url{https://huggingface.co/cross-encoder/ms-marco-MiniLM-L-12-v2} }. The total number of training steps is 50,000. For each step, we sample 40 queries and 10 hard negatives for each query. We set the learning rate to 2e-5 and the FLOPS $\lambda_d$ to 0.02. The max input length is set to 256. And the IDF values are calculated using the documents of MS MARCO dataset. The constant $S$ is set to 30 for scaling back the assembled scores.

\subsubsection{Indexing and Evaluation}
In this paper, a lexical search engine called OpenSearch\footnote{\url{https://opensearch.org/}}, is employed to construct the inverted index and perform the retrieval process as well. By leveraging the OpenSearch neural sparse feature, we can seamlessly integrate the writing and searching processes for custom learned sparse models. For evaluation metrics, we use the implementation of BEIR python toolkit to caculate the MRR, NDCG and recall rate. During evaluation, we use the IDF values derived from the MS MARCO dataset. The max input length is set to 512.

\subsection{Search Relevance Evaluation}
\subsubsection{In-domain Performance} As we fine-tune the model on MS MARCO dataset, we report the in-domain performance on this dataset.
Following \citet{formal2021splade}, we report the MRR@10 and Recall@1000 on MS MARCO dev set. We also report the NDCG@10 and Recall@1000 for the TREC DL 2019 evaluation set\footnote{TREC DL 2019 contains 43 queries for MS MARCO corpus annotated by human.}. We train the SPLADE-doc model using knowledge distillation techniques described by \citet{formal2021splade2} as baseline, and reference it as \textit{SPLADE-doc-distill} in other sections.
The results for baseline models are extracted from corresponding papers. The baseline models contain siamese dense encoders including ANCE~\cite{xiong2020approximate}, TCT-ColBERT~\cite{lin2021batch}, ColBERTv2~\cite{santhanam2022colbertv2}, RocketQA~\cite{qu2021rocketqa}, Roc\-ket\-QAv2~\cite{ren2021rocketqav2}, CoCondenser~\cite{gao2022unsupervised}, TAS-B~\cite{hofstatter2021efficiently}. And we also include sparse retrievers including BM25, SparTerm~\cite{bai2020sparterm}, DEEPCT~\cite{dai2020context}, doc2query-T5~\cite{nogueira2019doc2query} and SPLADE-series models~\cite{formal2021splade,formal2021splade2,formal2022distillation,lassance2024splade}. SPLADE-v3-Doc applies knowledge distillation on SPLADE-doc. It also employs tricks to improve the model's in-domain search relevance, and these tricks are orthogonal to what we proposed in this paper. The results are shown in Table \ref{tab:model_performance}.

\begin{table}[ht]
\centering
\caption{Evaluation result on MS MARCO dataset. Models marked with † means the model is trained and evaluated by us.}
\label{tab:model_performance}
\setlength{\tabcolsep}{2pt}
    \small
    \begin{tabular}{@{}lcccc@{}}
    \toprule
    \textbf{Model} & \multicolumn{2}{c}{\textbf{MS MARCO dev}} & \multicolumn{2}{c}{\textbf{TREC DL 2019}} \\
    \cmidrule(lr){2-3} \cmidrule(lr){4-5}
     & \textbf{M@10} & \textbf{R@1000} & \textbf{NDCG} & \textbf{R@1000} \\
    \midrule
    \multicolumn{5}{c}{\textit{Dense Retrievers}} \\
    \midrule
    ANCE   & 33.0 & 95.9 & 64.8 & - \\
    TCT-ColBERT  & 35.9 & 97.0 & 71.9 & 76.0 \\
    ColBERTv2  & 39.7 & 98.4 & - & - \\
    RocketQA  & 37.0 & 97.9 & - & - \\
    RocketQAv2  & 38.8 & 98.1 & - & - \\
    CoCondenser  & 38.2 & 98.4 & - & - \\
    TAS-B  & 34.7 & 97.8 & 71.7 & 84.3 \\
    \midrule
    \multicolumn{5}{c}{\textit{Sparse Retrievers}} \\
    \midrule
    SparTerm  & 27.9 & 92.5 & - & - \\
    DistilSPLADE-max  & 36.8 & 97.9 & 72.9 & 86.5 \\
    SPLADE-v3-DistilBERT  & 38.7 & - & 75.2 & - \\
    \midrule
    \multicolumn{5}{c}{\textit{Inference-free Sparse Retrievers}} \\
    \midrule
    BM25  & 18.4 & 85.3 & 50.6 & 74.5 \\
    DeepCT & 24.3 & 91.3 & 55.1 & 75.6 \\
    doc2query-T5  & 27.7 & 94.7 & 64.2 & \textbf{82.7} \\
    SPLADE-doc  & 32.2 & 94.6 & 66.7 & 94.7 \\
    SPLADE-doc-distill† & 36.5 & 96.9 & 69.8 & 74.2 \\
    SPLADE-v3-Doc & 37.8 & - & 71.5 & - \\
    Our Model†  & \textbf{37.8} & \textbf{97.5} & \textbf{72.1} & 79.8 \\
    \bottomrule
    \end{tabular}%
\end{table}

\begin{table*}[htbp]
\centering
\caption{Evaluation result on BEIR dataset. Models marked with † means the model is trained and evaluated by us. }
\label{tab:performance_comparison}
\small
\begin{tabular}{@{}lp{1.35cm}p{1.35cm}p{1.35cm}p{1.35cm}p{1.35cm}p{1.35cm}p{1.35cm}p{1.35cm}p{1.35cm}@{}}
\toprule
  & \multicolumn{4}{c}{\textbf{Inference-free Sparse Retriever}} & \multicolumn{2}{c}{\textbf{Sparse Retriever}} & \multicolumn{3}{c}{\textbf{Dense Retriever}} \\
  \cmidrule(lr){2-5} \cmidrule(lr){6-7} \cmidrule(lr){8-10}
Dataset & Our Model\textdagger & BM25 & SPLADE-doc-distill\textdagger & SPLADE-v3-Doc & SPLADE++\-SelfDistil & SPLADE-v3-Distil & ColBERTv2 & Contriever & TAS-B \\
\midrule
TREC-COVID        & 72.4 & 68.8 & 68.4 & 68.1 & 71.0 & 70.0 & \textbf{73.8} & 59.6 & 48.1   \\
NFCorpus          & \textbf{34.9} & 32.7 & 34.0 & 33.8 & 33.4 & 34.8 & 33.8 & 32.8 & 31.9   \\
NQ                & 53.1 & 32.6 & 48.8 & 52.1 & 52.1 & 54.9 & \textbf{56.2} & 49.8 & 46.3   \\
HotpotQA          & 67.9 & 60.2 & 62.6 & 66.9 & \textbf{68.4} & 67.8 & 66.7 & 63.8 & 58.4   \\
FiQA-2018         & \textbf{36.4} & 25.4 & 31.2 & 33.6 & 33.6 & 33.9 & 35.6 & 32.9 & 30.0   \\
ArguAna           & \textbf{49.1} & 47.2 & 37.7 & 46.7 & 47.9 & 48.4 & 46.3 & 44.6 & 42.9   \\
Touche-2020       & 28.7 & 34.7 & 25.6 & 27.0 & \textbf{36.4} & 30.1 & 26.3 & 23.0 & 16.2   \\
DBPedia-entity    & 40.5 & 28.7 & 35.9 & 36.1 & 43.5 & 42.6 & \textbf{44.6} & 41.3 & 38.4   \\
SCIDOCS           & \textbf{16.7} & 16.5 & 14.7 & 15.2 & 15.8 & 14.8 & 15.4 & 16.5 & 14.9  \\
FEVER             & 78.5 & 64.9 & 67.4 & 68.9 & 78.6 & \textbf{79.6} & 78.5 & 75.8 & 70.0  \\
Climate-FEVER     & 19.2 & 18.6 & 15.1 & 15.9 & 23.5 & 22.8 & 17.6 & \textbf{23.7} & 22.8 \\
SciFact           & \textbf{72.9} & 69.0 & 70.8 & 68.8 & 69.3 & 68.5 & 69.3 & 67.7 & 64.3   \\
Quora             & 84.2 & 78.9 & 73.0 & 77.5 & 83.8 & 81.7 & 85.2 & \textbf{86.5} & 83.5  \\
\midrule
Average           & \textbf{50.35} & 44.48 & 45.02 & 46.97 & \textbf{50.56} & 49.99 & \textbf{49.95} & 47.54 & 43.67 \\
\bottomrule
\end{tabular}
\end{table*}

The experiment results indicate that our model achieves SOTA in-domain performance among all inference-free retrievers. Additionally, the proposed model shrinks the performance gap between the inference-free retrievers and siamese sparse retrievers. Since there are trade-offs between search relevance and retrieval efficiency, we selected the Pareto optimal point based on the relationship curve between the two factors. It strikes a balance between search relevance and retrieval efficiency. Details can be found in Section \ref{sec:efficiency}.

\subsubsection{Out-of-Domain Performance on BEIR}
The purpose of out-of-domain(OOD) benchmarking is to test the model's generalization capacity in a zero-shot fashion, which represents a quantity of real production scenarios, especially for ones with limited resource for building a relevance tuning pipeline. Following the work of \citet{formal2022distillation,lassance2024splade}, we evaluate our model on a readily available subset of 13 datasets from the BEIR benchmark, excluding CQADupstack, BioASQ, Signal-1M, TREC-NEWS, and Robust04. The comparison is shown in Table \ref{tab:performance_comparison}.

 On the BEIR benchmark, our model's zero-shot search relevance substantially outperforms other inference-free sparse retrievers, surpassing SPLADE-v3-Doc by a significant margin of 3.3 NDCG@10 score. Our model's search relevance is comparable to SOTA siamese sparse retrievers and even outperforms strong siamese retrievers such as SPLADE-v3-DistilBERT and ColBERTv2. This result demonstrates that our model exhibits superior generalization capabilities. Moreover, we discovered that our model maintains stronger robustness in OOD settings compared to in-domain settings. 

\subsection{Retrieval Efficiency}
\subsubsection{Theoretical FLOPS number}
\label{sec:efficiency}
\begin{figure}[!t]
  \caption{Search relevance vs efficiency on BEIR for sparse retrievers. Our models are trained with different $\lambda_d$.}
  \includegraphics[width=\columnwidth]{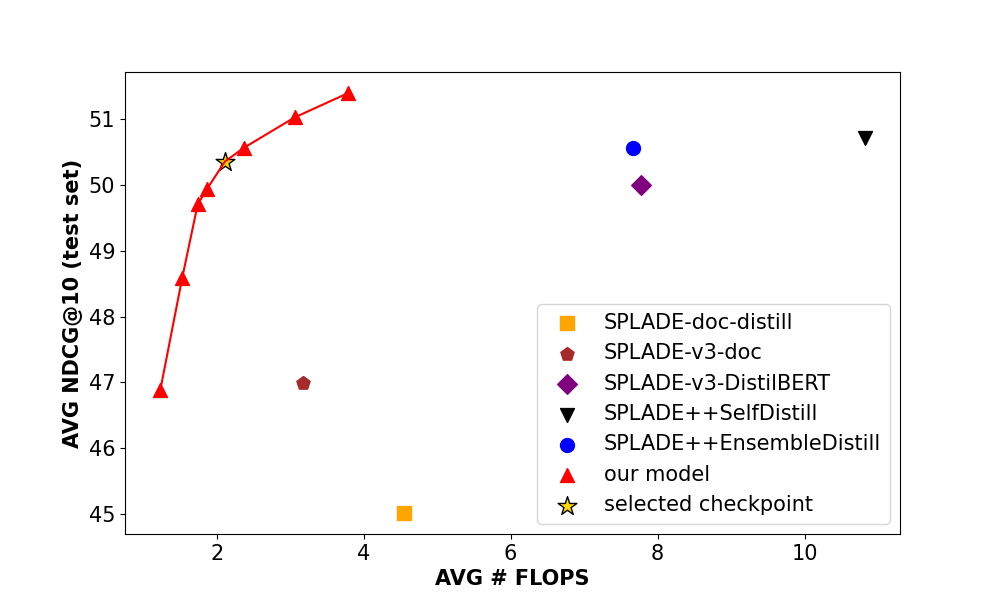}
  \label{fig:efficiency}
  \vspace{-20pt}
\end{figure}

FLOPS is the regularizer which controls the end-2-end efficiency of sparse retrieval. By adjusting $\lambda_d$ in equation \ref{eq:total_loss}, we can alter the degree of sparsity in the representation, as well as the computational cost incurred during the retrieval process ~\cite{formal2021splade, paria2020minimizing}. To investigate its impact, we conduct multiple sets of experiments by employing different values of $\lambda_d$ during fine-tuning, and measure the average FLOPS number for each query in BEIR datasets. The results are illustrated in Figure \ref{fig:efficiency}. The Pareto optimal point, which is regarded at the best trade-off between accuracy and efficiency, is marked in the diagram. In this paper, the corresponding checkpoint is used for other comparisons by default. Overall, in comparison to other inference-free sparse retrievers, our model exhibits not only superior search relevance but also enhanced retrieval efficiency. While siamese sparse retrievers possess search relevance comparable to or surpassing our model, our model's retrieval efficiency is significantly superior, which presents a substantial advantage in production settings.

\subsubsection{End-to-end search performance}
\label{sec:e2e-search}

To evaluate the efficiency of our model in real production settings, the benchmark is conducted on a distributed OpenSearch cluster. The end-to-end search performance is measured, where end-to-end refers to the process of sending a raw text search request and marking it complete upon receiving the search response. All workloads are included, such as tokenizer inference, network traffic etc. We compare our method with BM25 in terms of the 99th percentile (P99) search latency and average search throughput under different concurrency levels by adjusting the client number. By default, BM25 in OpenSearch employs heuristic optimizations, such as block-max WAND~\cite{ding2011faster}, while learned sparse retrievers do not have these optimizations.  We employ a simple heuristic optimization rule: first, we search for a preliminary result set using tokens with high IDF values, then we rerank the result set using all tokens\footnote{The implmentation is based on OpenSearch rescore query.}. The preliminary for this optimization rule is the involvement of IDF values. This optimization can boost search performance with negligible impact on search relevance. The results are listed in Table \ref{tab:e2e}. Our method achieves an efficiency very close to that of BM25. The heuristic optimization rule boost the search latency and throughput for our method about 10 percentage. When both employ heuristic optimizations, the latency of our method is approximately \textbf{1.1x} that of BM25.

\begin{table}[ht]
\centering
\caption{End-to-end search performance (milliseconds). Methods marked with † means the search process is optimized by heuristic rules.}
\label{tab:e2e}
\begin{tabular}{lcccccc}
\toprule
& \multicolumn{3}{c}{Client-side P99 latency} & \multicolumn{3}{c}{Mean throughput} \\
\cmidrule(lr){2-4} \cmidrule(lr){5-7}
Client \#& BM25† & Ours & Ours† & BM25† & Ours & Ours† \\ 
\midrule
5    & 13.4 & 21.7 & 17.6 & 784.2 & 484.8 & 586.2 \\
10   & 20.9 & 25.2 & 22.9 & 1150.9 & 910.4 & 1024.5 \\
20   & 35.4 & 38.2 & 38.7 & 1342.1 & 1183.4 & 1154.0 \\
40   & 56.7 & 66.2 & 62.3 & 1658.6 & 1460.5 & 1537.7 \\
80   & 74.7 & 91.7 & 81.1 & 2330.6 & 1858.19 & 2073.9 \\
\bottomrule
\end{tabular}
\end{table}

\subsection{IDF-Aware Penalty}
\subsubsection{Impact on search relevance}
\label{sec:idf-enhance}
To assess the impact of IDF-aware penalty on the zero-shot search relevance, we conduct an ablation study employing different IDF settings on BEIR benchmark. In the default setting, we use fixed IDF values derived from MS MARCO dataset. We also conduct experiments using IDF values from the corresponding BEIR datasets. With different settings, we examined the impact of (1) employing IDF-aware penalty in training phase (2) retrieval with IDF values derived from different sources. 
We conduct these experiments on our model and \textit{SPLADE-doc-distill}. The experiment results are shown in Table \ref{tab:idf}. From the experiment results, we obtain several conclusions: 
    (1) The IDF-aware penalty boost the model search relevance at large margin. Training and inference with IDF, both our model and SPLADE-doc-distill achieve much better search relevance.
    (2) For models without pre-training phase, using IDF derived from the test set has better search relevance compared with fixed IDF derived from training data. However, if the model has undergone extensive pre-training on large-scale data using the fixed IDF, the conclusion is the opposite.

\begin{table}[ht]
\caption{The search relevance impact of IDF on BEIR dataset.}
\label{tab:idf}
\small
\begin{tabular}{p{3cm}cccc}
\toprule
\multirow{2}{*}{Model}               & Trained & \multicolumn{2}{c}{Retrieval IDF} & \multirow{2}{*}{avg NDCG} \\ \cmidrule(lr){3-4} 
                                     & w/ IDF   & fixed    & BEIR    & \\
\midrule
Ours                            & \checkmark        & \checkmark              &             & \textbf{50.35}                                              \\
Ours                            & \checkmark        &                 & \checkmark           & 50.01                                               \\
Ours w/o IDF           & $\times$        &                 &             & 48.65                                               \\
SPLADE-doc-distill                 &  $\times$   &                 &             & 45.02                                               \\
SPLADE-doc-distill + IDF & \checkmark       &\checkmark        &             & 48.61                                               \\
SPLADE-doc-distill + IDF &\checkmark        &                 & \checkmark   & 49.13                                               \\

SPLADE-v3-doc   & $\times$     &                 &             & 46.97 \\                                             
\bottomrule
\end{tabular}
\end{table}

\subsubsection{Impact on retrieval efficiency and index size}
\begin{figure}[ht]
  \caption{The average FLOPS number, average token number per document and search relevance on BEIR datasets for models with and without IDF-aware penalty. The models are trained with different $\lambda_d$.}
  \includegraphics[width=\columnwidth]{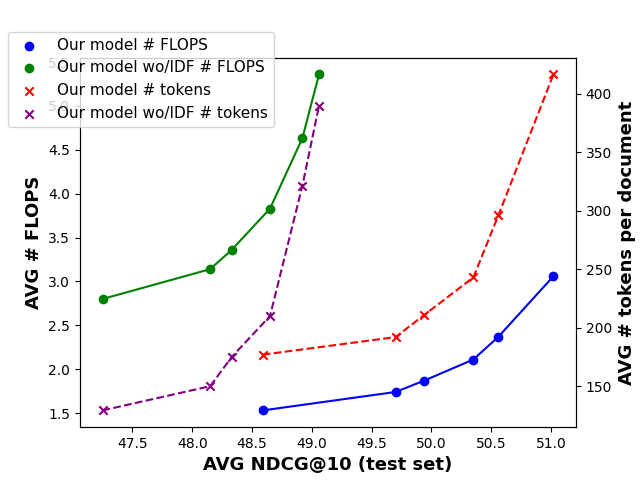}
  \label{fig:idf_effi}
  \vspace{-15pt}
\end{figure}

Experiments are conducted to measure the impact of IDF-aware penalty on retrieval efficiency. 
We conducted experiments to quantify the relationship between the expansion rate(average token number per expanded documents), retrieval efficiency(FLOPS number), and search relevance for our model. The comparison is made between our model with IDF-aware penalty and our model trained without IDF-aware penalty. The experimental results are depicted in Figure \ref{fig:idf_effi}. The experimental findings demonstrate that the IDF-aware penalty substantially augments the retrieval efficiency. For models with similar expansion rate, the FLOPS number of those trained with IDF-aware penalty is much smaller.

\subsection{Heterogeneous Knowledge Distillation}

We conducted an ablation study on the components of our proposed heterogeneous knowledge distillation to demonstrate their effectiveness. The detailed results are shown in Table~\ref{tab:pre-train}, yielding several notable findings:
\textbf{Knowledge distillation is a more effective optimization objective than the naive InfoNCE loss.} Utilizing supervision signals from one or more teacher models brings at least a 0.86 pts improvement (TAS-B) compared to model without pre-training, while the InfoNCE loss only improves 0.31 pts. \textbf{Assembling multiple teachers consistently show further improvements, no matter for dense teacher models (+0.44 pts) or sparse ones (+0.18 pts).} Our proposed heterogeneous knowledge distillation further enhances performance by 0.30 pts, amounting to a significant improvement of 1.74 pts. The process of normalizing outputs is crucial for model ensemble. The performance of using a simple additive ensemble method is close to that of a single teacher model (49.56). The significant performance drop empirically demonstrates the importance of addressing the scaling issue mentioned earlier.

\begin{table}[ht]
\caption{The search relevance impact of components in the pre-training stage.}
\label{tab:pre-train}
\begin{tabular}{lc}
\toprule
\textbf{Description}  & \textbf{BEIR avg score} \\
\midrule
\multicolumn{2}{c}{\textit{Pre-training techniques}} \\
\midrule
without pre-training & 48.61 \\
InfoNCE loss & 48.92 \\
ensemble KD, simply add & 49.56 \\
ensemble KD, norm and add (Ours) & \textbf{50.35} \\
\midrule
\multicolumn{2}{c}{\textit{Teacher models dense-only}} \\
\midrule
TAS-B & 49.47 \\
gte-large & 49.48 \\
TAS-B \& gte-large & 49.92 \\
\midrule
\multicolumn{2}{c}{\textit{Teacher models sparse-only}} \\
\midrule
opensearch-sparse & 49.66 \\
SPLADE++ED & 49.87 \\
opensearch-sparse \& SPLADE++ED & 50.05 \\
\bottomrule
\end{tabular}
\end{table}

\section{Conclusion}
In this paper, we proposed two novel approaches to significantly improve the search relevance of inference-free learned sparse retrievers while maintaining high efficiency. We introduced IDF-aware penalty to mitigate the uniform penalty on tokens, boosting both relevance and efficiency. We also developed a heterogeneous ensemble knowledge distillation framework leveraging strong dense and sparse retrievers for pre-training, enhancing generalization. Extensive experiments validate our methods' effectiveness. Our model achieves SOTA performance among inference-free sparse retrievers on BEIR, while maintaining end-to-end latency only 1.1x that of BM25.

\bibliographystyle{ACM-Reference-Format}
\bibliography{sample-base}
\end{document}